# On the Provenance of Pluto's Nitrogen (N$_2$)

**Kelsi N. Singer and S. Alan Stern**

*Southwest Research Institute, 1050 Walnut St. Suite 300, Boulder, CO 80302,USA*

ksinger@boulder.swri.edu

**Abstract**

N$_2$ is abundant in Pluto's atmosphere and on its surface, but the supply is depleted by prodigious atmospheric escape. We demonstrate that cometary impacts could not have delivered enough N mass to resupply Pluto's N$_2$ atmospheric escape over time; thus Pluto's N$_2$ is likely endogenous, and therefore was either acquired early in its history or created by chemistry inside/on Pluto. We find that cratering could excavate a considerable amount of N$_2$ to resupply the atmosphere against escape if the near-surface N$_2$ reservoir is deep. However, we find that this process likely falls short of that necessary to resupply the atmosphere against escape by at least an order of magnitude. We conclude that either the escape of N$_2$ from Pluto's atmosphere was on average much lower than the predictions for the current epoch, or that internal activity could be necessary to bring N$_2$ to the surface and resupply escape losses. Observations made by the New Horizons spacecraft in mid-2015 will yield further constraints on the provenance and evolution of Pluto's surface and atmospheric N$_2$, and could reveal evidence for past or present internal activity.

## 1. Introduction: Pluto's N$_2$ Atmospheric Loss Dilemma

In addition to Pluto's surface ice composition being dominated by molecular nitrogen (Cruikshank et al. 2015), Pluto's atmosphere consists of a >90% mole fraction of N$_2$ (Yelle & Elliot 1997), with surface pressures estimated on the order of ~10 μbars (Lellouch et al. 2009). Models predict the current N$_2$ escape rate is $10^{27}$–$10^{28}$ molecules s$^{-1}$, or 1.5x10$^{12-13}$ g yr$^{-1}$ (e.g., Erwin, Tucker, & Johnson 2013; Johnson et al. 2015; Tucker, Johnson, & Young 2015; Zhu, Strobel, & Erwin 2014). For comparison, the estimated global atmospheric mass of Pluto, based on pure N$_2$ of 10 μbars at 35 K, is ~3x10$^{16}$ g, demonstrating that the entire atmospheric N$_2$ reservoir will be lost in a tiny fraction of the age of the solar system. It is unknown if Pluto's atmosphere collapses over the course of Pluto's orbit (Hansen, Paige, & Young 2015; Olkin et al. 2015; and references therein). If the atmosphere exists for as little as 20% of Pluto's orbit, the escape rate would be reduced by a factor of 5. However, the higher solar ultraviolet fluxes in the past may have produced higher escape rates by a factor of several (Johnson, et al. 2015).

A linear extrapolation of the above escape rates indicates a total of $7 \times 10^{21-22}$ g $N_2$ has been lost over four billion years. This is equivalent to a condensed global $N_2$ surface layer on Pluto ~0.3-3 km in depth (Stern, Porter, & Zangari 2015b). Chondritic abundances of nitrogen (N) in all forms ($N_2$, $NH_3$, organics) are 3180 ppm or less (Lodders & Fegley 1998), with CI chondrites containing the highest values. Pearson et al. (2006) gives values for three CI chondrites that range between ~2000 and 5200 ppm; we adopt Lodders and Fegley (1998) as the middle of that range. Applying the same fractional N abundance (~0.0032) to the total mass of Pluto ($1.3 \times 10^{25}$ g) scaled by the rock mass fraction of Pluto of 0.65-0.7 (McKinnon 2015), one predicts a total Pluto N reservoir of $3 \times 10^{22}$ g. This N estimate is close to or lower than what is necessary to supply $N_2$ escape, if the escape rates have been similar to those stated above for the past four billion years. However, if Pluto's nitrogen abundance in all forms is instead on the one to several percent level (up to fractions of 0.02) as some sources suggest is the case with comets (e.g., Crovisier, Sylvio Ferraz, & Angel 2006; Jessberger et al. 1989; Mumma & Charnley 2011), and this N has been primarily converted to $N_2$ through chemical processes and given access to Pluto's surface, this could provide enough total $N_2$ mass to supply the atmosphere against escape.

Where did Pluto's $N_2$ originate from? To address this, we proceed as follows: in section 2 we evaluate whether $N_2$ delivered by comets to Pluto can resupply the nitrogen; in section 3 we explore potential excavation of endogenous $N_2$ by these same impactors. In section 4 we summarize our findings and briefly discuss the potential role of geologic activity for atmospheric resupply.

## 2. Can Delivery by Comets-Resupply the $N_2$?

*2.1 Predicted Impactor Populations*

Using observations of current KBO populations (Schlichting et al. 2012), typical KBO impact velocities onto Pluto of ~1-2 km s$^{-1}$ (e.g., Bierhaus & Dones 2015; Zahnle et al. 2003), and Pluto's cross section, one of us, Stern et al (2015a), estimated ~14,000 comets 1 km in diameter or larger would impact Pluto over 4 billion years (comparable to an earlier estimate by Durda and Stern [2000]). The rates derived in Zahnle et al. (2003) yield a similar order of magnitude estimate of ~11,200 impactors larger than 1 km in diameter over 4 billion years. Both Bierhaus and Dones (2015) and Greenstreet et al. (2015) conducted detailed analysis of KBO sub-populations and their estimates are 1840-5600 impactors > 1 km in diameter (nominal to maximum case), and 1400-8440, respectively. These estimates are based on what is known about the present-day impactor population and are taken to be valid for ~the past 4 billion years (starting after any Late Heavy Bombardment period and Charon's formation). Greenstreet et al. (2015) estimate a factor of a few is necessary to account for collisional and dynamical erosion of the Kuiper Belt over the past 4 billion years (e.g., Farinella, Davis, & Stern 2000; Stern & Colwell 1997; Weissman & Levison 1997). The impact flux model given by Bierhaus and Dones (2015) assumes the decay factor was outweighed by other uncertainties, and thus it is not considered in the model. Taking into account the maximum estimate of their model, it is possible that the uncertainties in the impactor flux could be up to an order of magnitude.

Adopting the Bierhaus and Dones (2015) estimate of impactor flux at Pluto (their Table 8, nominal case), a cumulative rate of a given size impactor and larger per year can be written as

$$rate(> d) = 4.6 x 10^{-7} d^{-1.837} \qquad (1)$$

where impactor diameter (*d*) is in km. The corresponding characteristic timescale between impacts of a certain size and greater, $\tau_{impact}$, is $1/rate(>d)$. To estimate the number of impactors as a function of size, we differentiate 1- *rate(>d)* to obtain:

$$spdf(d) = 8.4x10^{-7}d^{-2.837}. \quad (2)$$

Integrating this function between 1 m impactors and a hypothetical largest impact of ~60 km, we find that over 4 billion years there have been ~$6x10^8$ total impactors on Pluto, the majority of which would be in the smallest size range. Equation (2) further allows us to predict $1.2x10^5$ impactors between 100 m and 1 km in diameter over four billion years, with only ~1810 between 1 and 10 km, and 26 between 10 and 60 km, and 1 impactor that is 60 km or larger. Because we do not know the actual size of the largest impactor, for the following calculations we will assume the largest impactor in the past four billion yeas was 100 km as a reasonable upper limit. Multiplying equation (2) by the volume of a spherical impactor $(4/3)\pi(d*1000/2)^3$ and an estimate of overall comet density (1000 kg m$^{-3}$), and integrating from 1 m to 100 km impactors, yields a total mass of $3x10^{20}$ g delivered to Pluto. An independent calculation of total mass delivered by comets to Pluto by L. Dones (personal communication) found $3.7x10^{20}$ g, in good agreement with the estimate presented here.

*2.2 N Delivery by Comets*

Stern et al. (2015a) estimate an N$_2$ mass of $5x10^{10-11}$ g in a 1 km comet with a 50:50 ratio of volatiles to refractories, an overall density of ~1000 kg m$^{-3}$, and typical cometary N$_2$ volatile fractions of 0.002 to 0.0002 (Crovisier 1994; Rubin et al. 2015). However, given that N$_2$ is quite volatile, this specific molecule may be depleted in comets. In this paper we will consider the total nitrogen mass (i.e. N in all forms) per impactor by simply multiplying the volume of a spherical impactor $(4/3)\pi(d*1000/2)^3$, by the overall density (1000 kgm$^{-3}$), and the highest N fraction estimates (0.02), for an upper limit (Fig. 1a). We then estimate the mass of impactor-delivered N to Pluto per year (Fig. 1b), by multiplying equation (2) by the N mass per impactor:

$$M_{N\_delivered\_per\_year}(d) = 8.8x10^6 d^{0.163}, \quad (3)$$

where *M* and *d* are in g and km, respectively. Integrating equation (3) from 1 m up to an assumed largest impactor of 100 km (and multiplying by 4 billion years), would result in $6x10^{18}$ g N delivered to Pluto over the past four billion years.

From these considerations we conclude that the N mass delivered by comets is three-to-four orders of magnitude less than the 7x10$^{21-22}$ g N$_2$ lost by atmospheric escape over this same time period (the last ~4 billion years).  We conclude that even given uncertainties in the impactor flux (up to an order of magnitude), it does not appear that comets can deliver enough N to supply Pluto's N$_2$ atmosphere. The material influx onto Pluto during the first ~0.5 billion years of its history would contribute additional N, but this influx would need to be 3-4 orders of magnitude higher than that of the last 4 billion years, and remain near Pluto's surface, to supply the atmosphere at current escape rates.  Hence, except in the extremely improbable scenario that a recent, relatively large impact is supplying the current atmosphere, these calculations imply that either escape rate models overestimate total N$_2$ losses over time or endogenic sources must be responsible for replenishing Pluto's N$_2$.

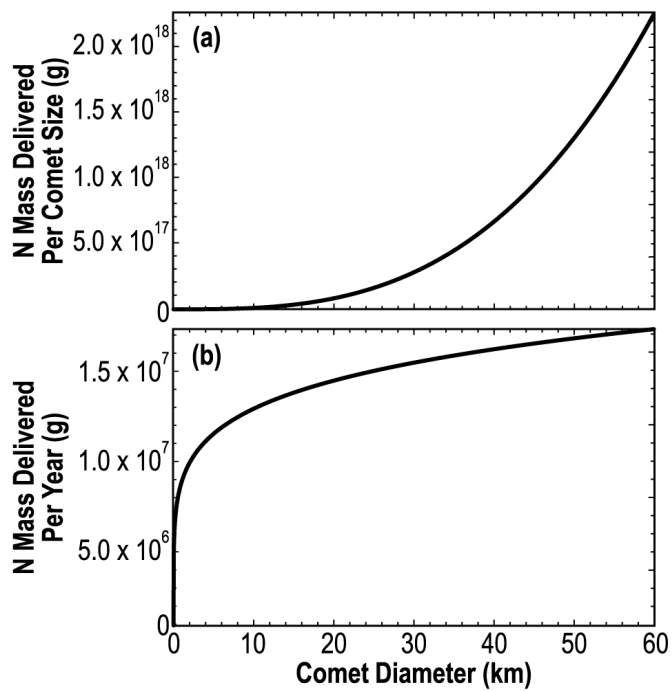

**Figure 1. N delivered by comets to Pluto. a.** The total delivered N per impactor size assuming an N fraction of (0.02), which gives an upper limit to what may be converted to N$_2$ on Pluto. **b.** Results from Equation (3) showing N delivered per year as a function of comet size. Integrating over comet size yields the total mass delivered per year.

## 3. Can Excavation by Cratering Resupply the N$_2$?

Cratering on Pluto may contribute to the resupply of atmospheric N$_2$ in two main ways: (1) cratering can excavate N$_2$ formerly thermally sequestered below inert lag deposits onto the surface as ejecta deposits, and (2) craters may directly (or indirectly) expose N$_2$ ice surfaces.

### 3.1 Lag deposits on Pluto

The surface and subsurface layers of Pluto likely contain a mixture of materials that are less volatile than N$_2$ (e.g., H$_2$O, CH$_4$, and other hydrocarbons). As N$_2$ sublimates from Pluto's surface, a lag deposit of these more refractory components is likely to build up and will tend to choke off Pluto's N$_2$ layer. The thickness $T_{lag}$ of a lag deposit can then be calculated by $T_{lag} = H_{sub} f_{invol} (\rho_{N2}/\rho_{invol})$, where $H_{sub}$ is the height of the sublimated layer, $f_{invol}$ is the involatile fraction of the surface (involatile/N$_2$), which we assume for simplicity to be identical everywhere, and ($\rho_{N2}/\rho_{invol}$) is the density ratio of the volatile to involatile materials. The distribution of materials of the surface and near-surface layers is not well known, thus we present a range of scenarios for $f_{invol}$ in Fig. 2.  It is unknown at this time if this N$_2$ ice detected on Pluto's surface is a thin veneer, or a deeper layer, or how thick of lag deposits may have built up over time. We take ($\rho_{N2}/\rho_{invol}$) of ~one for N$_2$ mixed with involatile ices, and ~1/3 for N$_2$ with a rocky/organic component.

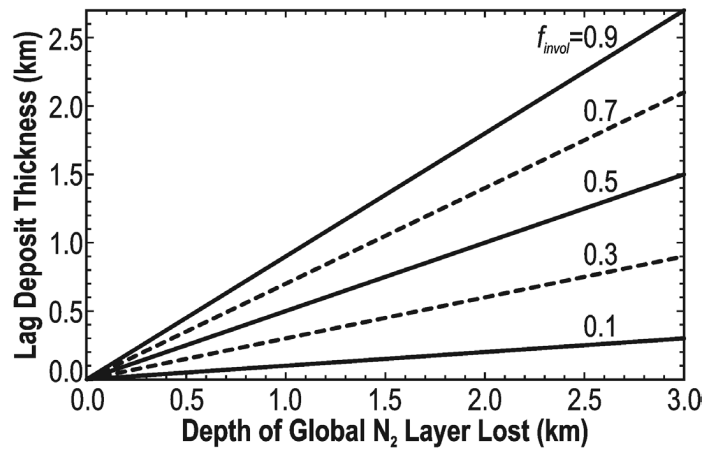

**Figure 2.** Lag deposit thicknesses as a function of the depth of the sublimated N$_2$ layer ($H_{sub}$) and for varying involatile/N$_2$ fractions for the surface and near-surface layers.

We thus calculate that if ~0.3–3 km of N$_2$ have been lost to escape (the thickness of a globally averaged surface layer), this loss could leave a total lag deposit built up over 4 billion years of ~0.15–1.5 km (Fig. 2) for an involatile fraction of 0.5.  Although this estimate is crude, it serves for subsequent analysis we will use to explore how lag deposits would affect N$_2$ resupply against escape.  We expect such a lag deposit to lie beneath a thin veneer of N$_2$ deposited by the most recent seasonal atmospheric deposition cycle (Stern, Trafton, & Gladstone 1988).  Pluto's complex seasonal cycles (Young 2013), potential polar caps (Olkin, et al. 2015), and varied surface albedo patterns (Buie et al. 2010) suggest the surface ices are heterogeneously distributed. If the $N_2$ must be supplied by a smaller percentage of Pluto's surface area then sublimation would need to proceed to larger depths to supply the same $N_2$ mass, hastening the development of lag deposits.

*3.2 $N_2$ excavation by cratering*

The $N_2$ mass excavated by impacts reaching below Pluto's lag deposit would be available to resupply the atmosphere. For simple craters on Pluto (less than 4 km in diameter; Moore et al. 2015) the final crater diameter, $D_f$, is taken as $1.2*D_{tr}$ based on lunar data (Pike 1977). For the final diameter of complex craters we use McKinnon and Schenk (1995; their equation 1). Using this, a 5 km impactor makes an ~32 km final crater (scaling explained below), and excavates to a depth of ~2 km. This crater excavates ~3 x $10^{17}$ g of material, which is similar to Pluto's current atmospheric mass (see Table 1 for additional examples).

**Table 1. $N_2$ Excavated by Cratering**

| Impactor Diameter [km] | Transient Crater Diameter[a] [km] | ~Final Crater Diameter[b] [km] | ~Excavation Depth[c] [km] | Excavated $N_2$ Mass[d] [g] | $\tau_{impact}$[e] [yr] | Number this size and larger in 4 Ga[e] |
|---|---|---|---|---|---|---|
| 0.01 | 0.15 | 0.18 | 0.015 | 1.4x$10^{11}$ | 4.6x$10^2$ | 8.6x$10^6$ |
| 0.05 | 0.54 | 0.65 | 0.05 | 6.0x$10^{12}$ | 8.9x$10^3$ | 4.5x$10^5$ |
| 0.1 | 0.92 | 1.1 | 0.09 | 3.1x$10^{13}$ | 3.2x$10^4$ | 1.3x$10^5$ |
| 0.5 | 3.3 | 3.9 | 0.33 | 1.4x$10^{15}$ | 6.1x$10^5$ | 6.6x$10^3$ |
| 1 | 5.6 | 6.8 | 0.56 | 6.9x$10^{15}$ | 2.2x$10^6$ | 1.8x$10^3$ |
| 5 | 20 | 32 | 2 | 3.1x$10^{17}$ | 4.2x$10^7$ | 95 |
| 10 | 34 | 59 | 3.4 | 1.6x$10^{18}$ | 1.5x$10^8$ | 26 |
| 20 | 59 | 108 | 5.9 | 8.0x$10^{18}$ | 5.3x$10^8$ | 8 |
| 40 | 102 | 197 | 10 | 4.1x$10^{19}$ | 1.9x$10^9$ | 2 |
| 60 | 140 | 280 | 14 | 1.1x$10^{20}$ | 4.0x$10^9$ | 1 |

[a]Scaling from equation (4). [b]Conversion from transient to final diameter described in section 3.2. [c]Excavation depth taken as ~1/10$^{th}$ the transient crater diameter (Melosh 1989). [d]See equation (5). [e]Mean time between impacts and number in four billion years (for the given size and larger) from Bierhaus & Dones (2015), nominal rates from their Table 8.

To calculate the total $N_2$ excavated by cratering on Pluto, we use equation (2), which gives the cratering rates per year, multiply it by an estimate of the $N_2$ excavated by a given size impact (derivation below), sum over all impactor sizes, and multiply by four billion years. This involves scaling from the impactor size to the crater size (e.g., Holsapple 1993). Little data exists on the mechanical behavior of $N_2$ ice, so we follow the example of previous authors in using $H_2O$ ice as an order of magnitude estimate (e.g., Bierhaus & Dones 2015; Stern, et al. 2015b). $N_2$ ice is a weaker material, thus in theory craters would scale differently in the strength regime. This would only affect craters on Pluto smaller than a few hundred meters, which are created by impactors smaller than a few tens of meters. For larger impacts in the gravity regime, the density contrast between the impactor and target plays a role, but the density of $N_2$ is similar to that of $H_2 0$ ice at Pluto temperatures (Scott 1976). The scaling formula derived for primary craters (depth/diameter = 0.2) for cometary impactors into water ice in the gravity regime with a 45° impactor (details in Singer et al., 2015) is

$$D_{tr}(d) = 0.86 d^{0.783}(g/U^2)^{0.217} \quad (4)$$

where $D_{tr}$ is the transient crater diameter, $g$ is surface gravity (0.66 m s$^{-2}$ on Pluto), $U$ is the impact velocity (we use 2 km s$^{-1}$ as a typical speed), and all variables are in MKS units. Here we use a paraboloid of revolution as the crater shape, where the volume is $(1/2)\pi R_{tr}^2 H_{tr}$. $R_{tr}$ and $H_{tr}$ are the transient crater radius and depth, respectively, and both can be written in terms of $D_{tr}$ and thus the impactor diameter ($d$) through equation (4).

Using the scaling in equation (4), half the transient crater volume as the approximate amount of excavated material, $N_2$ ice density of 1000 kg m$^{-3}$ (Scott 1976), and for an upper limit assuming an 100% $N_2$ ice layer, we arrive at the mass in grams excavated per impact size (Fig. 3a) of

$$M_{N2_{excavated}}(d) = 7.0 \times 10^{15} d^{2.353} \quad (5)$$

where again $d$ is in km. Multiplying equation (5) by equation (2), we derive an equation for the $N_2$ mass (g) excavated per year (Fig. 3b) as

$$M_{N2\_excavated\_per\_year}(d) = 5.9 \times 10^9 d^{-0.484}. \quad (6)$$

The negative exponent on impactor diameter in equation (6) reflects the fact that both the cratering efficiency and the number of impacts decreases as impactor size increases. Integrating from 0 to 100 km for impactor diameters results in a total of $5 \times 10^{20}$ g $N_2$ excavated over four billion years. The

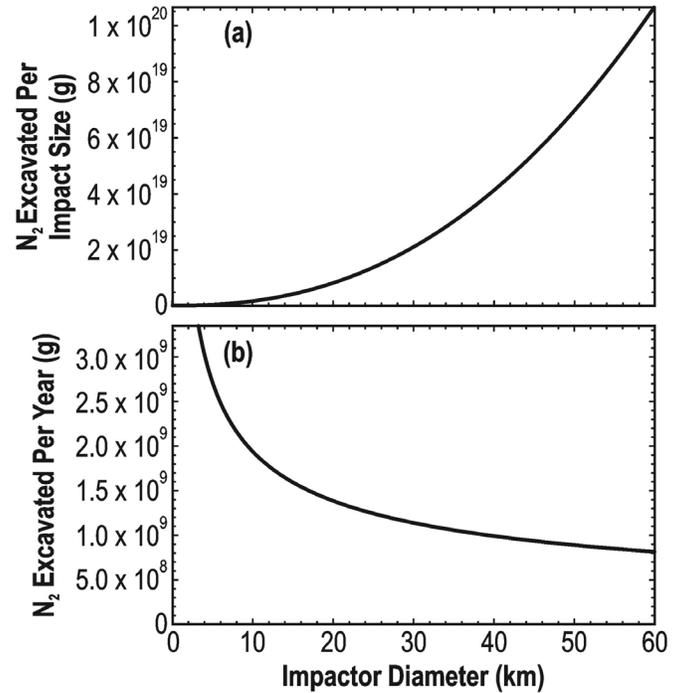

**Figure 3. $N_2$ excavated by cratering. a.** Mass of $N_2$ excavated (g) per impact size assuming a near surface layer of solid $N_2$ at least as thick as the excavation depth of the impact (up to 14 km deep for the largest impactors/craters) as given by equation (5). **b.** The result of evaluating equation (6) showing $N_2$ excavated per year as a function of impactor size. Integrating over impactor size yields the total mass delivered per year. See Table 1 for conversion of impactor size to crater size and approximate excavation depth.

low primary impact speeds on Pluto mean only a small fraction of the ejecta, less than 1% of the impactor mass, will have velocities higher than Pluto's escape speed, and not much material is vaporized (Bierhaus & Dones 2013),

Even with uncertainties in the impactor flux, and the extremely conservative assumption that all excavated material is pure $N_2$, excavation from cratering does not seem to be a likely source for supplying *all* of Pluto's atmospheric mass loss over time. In fact, it apparently falls short by an order of magnitude or more[1].

---

[1]This result is further strengthened by noting that only craters that are deep enough to punch through the lag can excavate $N_2$. Using 0.15–1.5 km as a range of possible lag deposit thicknesses means craters smaller than ~1.5–15 km (corresponding to impactors of ~0.2–3.5 km) will not punch through. If we take a second integration of equation (6), this time from 3.5 to 100 km impactors, the mass excavated only changes by a small fraction to $4.0 \times 10^{20}$ g $N_2$ over 4 billion years, as the larger craters play a more significant role in this calculation and the effect of a lag deposit is second order.

*3.3 Sublimation of $N_2$ through crater floors and walls*

    Cratering can also contribute new $N_2$ to atmospheric resupply by exposing buried $N_2$ next to crater walls/floors and as the thermal wave from solar insolation drives subsurface $N_2$ sublimation. We use 2 times the annual thermal skin depth as a rough upper limit length scale over which this sublimation is likely to occur. The thermal skin depth, or $L_T$, is given by $\sqrt{(\kappa\tau)/(\pi\rho c_p)}$, where $\kappa$ is thermal conductivity, $\tau$ is the thermal variation timescale (conservatively we adopt the annual solar insolation on Pluto of 248 years), $\rho$ is material density, and $c_p$ is heat capacity. For a pure slab of $N_2$ ice, we estimate that $L_T$, or one e-folding depth of the annual wave would be ~20 m.

    The $N_2$ mass sublimated from the ice near a crater cavity is estimated by taking the volume of a paraboloidal shell with an inner radius of $D_f/2$ and an outer radius of $D_f/2 + 2L_T$, where $2 L_T$ is 0.04 km in this case (note the use of final, rather than transient, diameter here). Multiplying the volume of the paraboloidal shell by equation (2) gives

$$M_{N2\_sublimated\_per\_year}(d) = 2.0x10^{10}(0.04+d)^{2.353}d^{-2.837} - 2.0x10^{10}d^{-0.484} \quad (7)$$

($d$ in km), and integrating over all 1-m to 100-km-diameter impactors results in $8x10^{21}$ g $N_2$ lost through crater wall and floor sublimation over four billion years, assuming pure $N_2$ under the cavity of each crater. In this simple upper limit calculation, the sublimation contribution is ~one order of magnitude higher than that excavated by cratering due to the fact that the amount excavated by small craters (of which there are many) is less than the amount sublimated. This simple estimate does not attempt to take into account the actual composition of subsurface materials, any lag deposits (discussed below), the actual final shape of large versus small craters, or any effects of the varying insolation that would be received by a crater cavity at different latitudes on Pluto (there are no crater shadows in this model). These additional effects would work to reduce the amount of $N_2$ sublimated. It is possible that impact heating or other effects could enhance sublimation temporarily at the impact site, but we consider this is to be a second order effect because of the low impact velocities found at Pluto.

    If a lag deposit is present, smaller craters would not expose fresh $N_2$ surfaces. To calculate the effect of a lag deposit, we assume that the excavation depth plus 40 m ($2L_T$ for pure $N_2$) must be greater than the lag deposit thickness. For the lower limit lag deposit of 0.15 km, craters smaller than ~1.3 km in final diameter will not punch through. Integrating over impactors from 125 m to 100 km, the material sublimated is considerably reduced, down to $4x10^{19}$ g $N_2$ sublimated over four billion years (for a pure $N_2$ surface). For a 1.5 km lag deposit, again integrating from 3.5 to 100 km in impactor diameter, the mass sublimated is reduced to $5x10^{17}$ g $N_2$ over four billion years.

    All of these various estimates of exposed $N_2$ based on the presence of a lag deposit fall short by two-to-five orders of magnitude of the amount necessary to resupply atmospheric escape.

## 4. Summary and Evidence for the Possibility of Internal Activity on Pluto and $N_2$

    Based on the first-order analysis conducted here, it is does not seem that either cometary import or cratering-related excavation/sublimation effects can resupply Pluto's atmospheric $N_2$ escape losses. Given the also demonstrated difficulty of delivering enough $N_2$ with comets, we suggest that either atmospheric escape rates have been overestimated or cryovolcanism or another tectonic or geodynamic means of $N_2$ resupply may be necessary to resupply Pluto's atmosphere against escape and the buildup of an involatile lag deposit resulting from the escape

process. The volumetric resupply rate for Pluto's atmosphere is 0.0015-0.015 km$^3$ yr$^{-1}$. This value is ~30-13,000 times lower than Earth's resurfacing rates of ~0.4-20 km$^3$ yr$^{-1}$, where the range is for intraplate volcanism versus total intrusive plus extrusive volcanism (Crumpler et al. 1997; Smrekar, Stofan, & Meuller 2014; Strom, Schaber, & Dawsow 1994). The higher Pluto rate is similar to estimated resurfacing rates for Triton of 0.01 km$^3$ yr$^{-1}$ (Stern & McKinnon 2000), implying this rate of activity on Pluto is possible. Evidence of subsurface activity on Pluto could reveal itself in images returned by New Horizons.

## 5. Acknowledgements

We thank an anonymous reviewer for helpful comments that improved and clarified this manuscript. We also thank Luke Dones, Randy Gladstone, and Eliot Young for helpful discussion on this manuscript. This work was supported by NASA's New Horizons mission contract NASW-02008 to the Southwest Research Institute.